# Characteristic Direction Approach to Identify Differentially Expressed Genes

Neil R. Clark[1,*], Kevin Hu[1], Edward Y. Chen[1], Qioanan Duan[1], Avi Ma'ayan[1,*]

[1]*Department of Pharmacology and Systems Therapeutics, Systems Biology Center New York (SBCNY), Icahn School of Medicine at Mount Sinai School, New York, NY 10029 USA*

*Corresponding authors: avi.maayan@mssm.edu and neil.clark@mssm.edu

**Abstract**

Genome-wide gene expression profiles, as measured with microarrays or RNA-Seq experiments, have revolutionized biological and biomedical research by providing a quantitative measure of the entire mRNA transcriptome. Typically, researchers set up experiments where control samples are compared to a treatment condition, and using the t-test they identify differentially expressed genes upon which further analysis and ultimately biological discovery from such experiments is based. Here we describe an alternative geometrical approach to identify differentially expressed genes. We show that this alternative method, called the Characteristic Direction, is capable of identifying more relevant genes. We evaluate our approach in three case studies. In the first two, we match transcription factor targets determined by ChIP-seq profiling with differentially expressed genes after the same transcription factor knockdown or over-expression in mammalian cells. In the third case study, we evaluate the quality of enriched terms when comparing normal epithelial cells with cancer stem cells. In conclusion, we demonstrate that the Characteristic Direction approach is much better in calling the significantly differentially expressed genes and should replace the widely currently in used t-test method for this purpose.

Implementations of the method in MATLAB, Python and Mathematica are available at: http://www.maayanlab.net/CD.

# Introduction

The ability to measure mRNA levels at a genomic scale allows the development of clinical markers for disease, reveals the heterogeneity of histologically identical cancers, and sheds light on diverse biological mechanisms. After estimating the relative or absolute expression level of all transcripts, testing of statistical hypotheses follows [1]. It is this step of statistical inference that reveals biological insights. Typically, the statistical hypotheses tested are concerned with the difference in gene expression levels between two biological conditions, for example, normal vs. diseased tissue, or perturbed vs. unperturbed cells. One common approach, which remains popular due to its simplicity, is the fold-change. However, this approach does not incorporate the variance, which arises from biological and experimental sources, and hence the fold-change approach does not offer any estimate of confidence [2, 3]. Because of this, the fold-change method is now regarded as an insufficient statistic for identifying differentially expressed genes [3, 4]. In most studies, the t-test and its extension, analysis of variance (ANOVA), is applied to identify differentially expressed genes between two or more groups.

Few attempts have been made to improve upon the popular t-test. One development was the realization that variance shrinkage improves statistical power. Variance shrinkage attempts to tackle the problem of typically having only few samples measuring a large number of genes [5-7]. In addition, methods that identify differentially expressed gene-sets instead of single genes have been developed [8-13]. These approaches attempt to facilitate the problem of biological interpretation, which can be challenging when faced with a long list of differentially expressed genes [9, 14], while also increasing statistical power. Typically, the differentially expressed gene-sets are predefined, derived from resources such as the gene ontology (GO) [14].

Here we demonstrate a different approach, called the Characteristic Direction, which identifies differentially expressed genes geometrically. The approach attempts to make maximal use of the information in the preprocessed expression data by using linear classification approaches which take into account the variances and the correlations of the genes in a shared manner. The Characteristic Direction approach provides an intuitive visualization of the differential expression. Through a couple of case studies we demonstrate that the Characteristic Direction approach outperforms other commonly applied methods.

# Methods

*The Characteristic Direction*

We aim to extract the genes which are most important in distinguishing between two biological conditions as represented by their expression profiles. To do this we consider the logarithmic gene expression space where the logarithm of the estimator of expression of each gene is regarded as a coordinate; any gene expression profile can be represented as a point in this space. We consider the logarithm because distances in this space correspond to fold changes rather than absolute changes. The transcriptional profile of any given biological condition can be represented as a joint probability distribution defined through this expression space; the localization and structure of this distribution captures both the overall pattern of gene expression as well as the natural biological variability and gene-gene correlations. We illustrate this idea in the case where there are only two genes for the purpose of intuitive visualization. However, the typical space will have thousands of dimensions. Expression profiles from two different biological conditions are represented by red and blue points (Fig. 1). If there is a difference between the biological conditions in terms of gene expression, then the distribution of these points will occupy different regions in the two dimensional expression-space. We shall restrict our attention to cases in which the difference between the two conditions can be characterized by a single direction in the space. This direction may be identified with the perpendicular to the classification boundary between the two conditions (Fig. 1A). The direction can be represented by a unit vector, and then the relative magnitudes of the components of this vector correspond to the relative significance of the genes in distinguishing between the two biological conditions. In fact, because the direction is represented as a unit vector, the sum of the squares of the components is equal to unity, and we can interpret the square of each component as being equal to the fractional contribution of the corresponding gene to the total difference between the transcriptional profiles of the two biological conditions under comparison. This can be used to rank the genes and isolate the most relevant genes. The direction identified, maximally distinguishes between the two conditions; this can be seen by projecting the data onto this direction (see the inset distributions in the upper right corners of Fig. 1A-C). In some cases this is not necessary because the differentially expressed genes will be identified regardless of the axis rotation (Fig. 1A). However, in many situations the distinction between expression profiles may not be well characterized by the standard t-test approach for identifying differentially expressed genes (Fig. 1B). Only after projecting the profiles onto a new and more relevant direction, the direction perpendicular to the classification boundary (dashed line), do we clearly see the differences between the two biological conditions. In this example (Fig. 1B), both genes are important in distinguishing the difference between the two conditions; however, neither gene would be detected as differentially expressed with a gene-by-gene test of difference of location such as a t-test. When the difference between two conditions is slight, no single gene would be detected as differentially expressed (Fig. 1C). However, after projecting the data onto the direction perpendicular to the linear classification boundary (upper right inset), we maximize our chances of detecting differentially expressed genes. The direction of the perpendicular then provides an estimate of the relative importance of each gene in the subtle distinction. Even a case where many genes will be detected as differentially expressed the direction perpendicular to the classification boundary is still relevant as this approach takes the variance and correlation structure of the distribution into account (Fig. 1D).

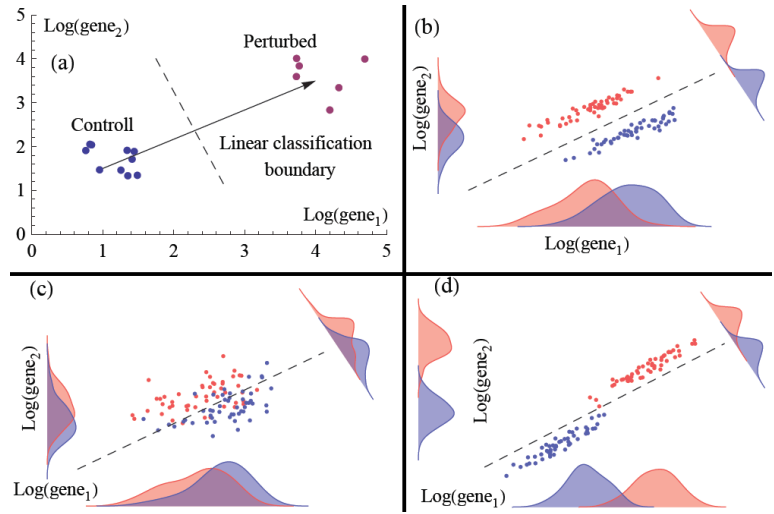

**Fig 1.** Toy example to demonstrate the concept of the Characteristic Direction identified with a hyperplane which defines the boundary of a linear classification between two sets of gene expression profiles. a) The boundary in gene expression space is represented by a dashed line, whereas the perpendicular direction is indicated by an arrow. b-d) Illustrative examples of the distribution of expression profiles in two-dimensional logarithmic gene expression space, along with the distributions as projected onto the gene axes and the direction perpendicular to the classification boundary hyperplane.

Having identified the characteristic direction and interpreted the squared coefficients as the fractional contribution to the total difference between the transcriptional profiles of the two biological conditions being compared for each gene, we may examine the significance of those differentially expressed genes. This can be achieved by using the squared coefficients to rank the genes and then evaluate enrichment of gene sets in the extreme ranking. However, this approach involves an arbitrary cut-off. An alternative method, consistent with the geometrical approach, presents itself naturally here: Every set of genes identifies a sub-space in expression space which consists of the span of each gene in the set. The principal angle [15] between the Characteristic Direction and this subspace quantifies the alignment of the direction characterizing the difference between the biological states with the genes in the list. Principal angles are related to the canonical correlations discovered in the 1930's [16] which have found many applications in statistical analysis in economics and meteorology, but are largely overlooked in the field of statistical learning and systems biology. We can also use the null-distribution of the isotropic principal angles to define a p-value that is normalized across gene sets of different length. We also adopt this approach when we perform gene-list enrichment analysis to validate the Characteristic Direction method.

*Approaches to compute the Characteristic Direction*

We are interested in the cases in which a single direction can characterize the difference in expression between two sets of samples. Linear classification methods divide a space with hyperplanes which have orientations defined by a Characteristic Direction. These methods include classification by linear regression (LR1), logistic regression (LR2), and linear discriminant analysis (LDA). LR1 classifies by minimizing a sum-of-squares loss function. LR2 explicitly models the log-ratios of the conditional classification probabilities as linear functions. LDA employs an application of Bayes' theorem to the class posteriors and assumes that the classes have equal covariance matrices. This appears to be a strong assumption. However, LDA produces similar results to LR1 and LR2 in the context of our analyses. Each of the above methods provides a means of deriving the Characteristic Direction that we require. Here we test these methods as well as derive an alternative non-parametric approach to arriving at a comparable Characteristic Direction. In this non-parametric approach we permute the class labels, destroying the information about the classes while preserving the correlation structure of the genes, and then evaluate the difference between the centroids. In this way we derive a null set of direction vectors to which we can compare the actual difference between centroids. The correlation structure of the null set is used to represent all the vectors in the principal component space. The components of the actual difference between the centroids are then scaled by the standard deviation of the null distribution to extract the most significant direction. Expressed in the full logarithmic gene expression space this approach provides the desired Characteristic Direction. We note that all the above methods pool the information from all genes, and extract information from the correlation structure to characterize the difference between the expression profiles. This should increase the statistical power and lead to more robust and reliable result that is most relevant to the characterization of the difference between the relevant biological conditions under investigation.

*The linear regression approach (LR1)*

The first approach uses linear regression of an indicator matrix to derive a hyperplane classification boundary. First, because data with a sample size $N$ spans a subspace of dimension $N$ in the expression space, we use principal component analysis (PCA) to derive a coordinate system for this subspace. For numerical efficiency we reduce the dimensionality, retaining enough principal components to capture a fraction $1 - \epsilon$ of the variance ($\varepsilon = 10^{-3}$) and when $N$ is large (>20) we limit the number of components. In this subspace an indicator variable is used to describe the class of each sample, for example, normal or disease conditions. For example, with two classes, the indicator $Y_k$ is where $k = 1$ or $k = 2$ for the first and second classes. These are collected into a vector and the vector corresponding to each of the $N$ samples is collected into a $N \times 2$ matrix. With the $N \times (p + 1)$ expression matrix $X$, where $p$ is the number of genes, i.e., probe sets on an array chip. We can then solve the linear regression problem as follows:

$$\hat{Y} = X(X^T X)^{-1} X^T Y \qquad (1)$$

where $\hat{Y}$ is the estimated class from the linear regression fit. This provides a linear relationship between the input data $X$ and the indicator variable $Y$, via the $(p + 1) \times 2$ coefficient matrix,

$$B = X(X^T X)^{-1} X^T \qquad (2)$$

This coefficient matrix defines a hyperplane, described by the solution to the equation $x.B = 0$. The unit normal vector $\hat{b}$ to this hyperplane is identified with the direction characteristic of the differential expression between the two samples. The direction, derived in principal components, is coordinate transformed into the full expression space. The Logistic Regression (LR2) and LDA produce similar results to the Linear Regression (LR1) approach and are not described in detail for conciseness.

*The non-parametric approach (NP1)*

We developed an alternative approach to identifying the Characteristic Direction in which we use sample permutations to calculate a null-distribution of directions. The null distribution is generated by permuting the sample labels but not the gene labels in order to preserve the global correlations. Then the difference between the centroids is evaluated as follows:

$$d = \overline{X_2} - \overline{X_1} \qquad (3)$$

where $\overline{X_1}$, and $\overline{X_2}$, are the permuted expression data matrices for class 1 and 2. This process is repeated many times, at least 100, to obtain the null distribution of the characteristic directions. We then correct the observed difference between centroids by comparing it to this null distribution. This is again achieved with the linear regression classification boundary as described above.

*Visualization of expression levels in the characteristic direction*

The directions we calculate are intended to be characteristic of the difference between the two sets of samples. In order to visualize the extent to which the directions do this, we project the data onto them and plot the distribution. If the data matrix, $X$, has dimensions $(n, m)$ where $n$ is the number of genes and $m$ is the number of samples, and the $n$-dimensional unit vector column, $\hat{d}$, is parallel to the characteristic direction, then the projected data vector is given by:

$$X_p = \hat{d}.X \qquad (4)$$

The components of this vector correspond to the projection of each microarray sample onto the Characteristic Direction. We can take this a step further by successively projecting to calculate a hierarchy of Characteristic Directions. This is done by iteratively projecting the data onto the classification hyperplane, defined by the orientation vector $\hat{b}_i$, after iteration $i$,

$$X_{i+1} = X_i - \hat{b}_i(\hat{b}_i.X_i) \qquad (5)$$

Then by taking each $X_i$ and projecting onto its corresponding characteristic direction, $\hat{b}_i$,

$$X_{proj,i} = \hat{b}_i.X_i \qquad (6)$$

We can then plot the data as projected onto a subspace of expression space which has a dimension equal to the number of iterations. The directions we calculate are intended to be characteristic of the difference between the two sets of samples.

*Calling significant genes using the characteristic direction*

In the case that the difference between the samples can be characterized by a single vector, the magnitude of the vector quantifies the magnitude of the difference in expression between the two groups of samples. However the unit vector parallel to the Characteristic Direction, quantifies the contribution of each gene to this total difference because,

$$\sum_{i=1}^{i=n} \hat{b}_i^2 = 1 \qquad (7)$$

The contribution of an individual gene to the total difference between the sample classes is interpreted as $\hat{b}_i^2$. The contribution of some subset, $S$, of genes to the total difference between the microarray sample classes, the two biological conditions, is:

$$\sum_{i \in S} \hat{b}_i^2 = \alpha \qquad (8)$$

*Calling differentially expressed genes with the t-test for comparison*

For comparison we employ the commonly used Welsh t-test to identify differentially expressed genes. Those genes which are significant at a given false discovery rate (FDR) determined after applying Benjamini-Hochberg multiple hypothesis test correction are selected for enrichment analysis. With sample means $\bar{X}_i$, variance $\sigma_i^2$, and size $N_i$, the test statistic is given by:

$$t = \frac{X_1 - X_2}{\sqrt{\frac{\sigma_1^2}{N_1} - \frac{\sigma_2^2}{N_2}}} \qquad (9)$$

With the degrees of freedom given by,

$$df = \frac{\left(\frac{\sigma_1^2}{N_1} - \frac{\sigma_2^2}{N_2}\right)^2}{\frac{\sigma_1^4}{N_1^2(N_1-1)} + \frac{\sigma_2^4}{N_2^2(N_2-1)}} \qquad (10)$$

The p-value is then derived from the cumulative distribution function of the Student t distribution where p-values are corrected for multiple hypotheses testing with the Benjamini-Hochberg test, resulting in equivalent q-values.

*Calculation of the null-distribution of principal angles*

We calculate an analytical expression which provides the null-distribution of the first principal angle between a line and an n-dimensional subspace. This is used to generate a p-value for the enrichment of gene sets. We begin by calculating the distribution of the angles between a pair of isotropic directions. The surface area of an n-sphere is given by:

$$S_n(r) = \frac{2\pi^{\frac{n+1}{2}}}{\Gamma\left(\frac{n+1}{2}\right)} \qquad (11)$$

The probability distribution of the angle between two isotropic directions in an n-dimensional space is given by the ratio:

$$\begin{aligned} pdf(\theta) &= \frac{S_{n-2}(Sin(\theta))}{S_{n-1}(1)} \\ &= \pi^{-\frac{1}{2}} \frac{\Gamma\left(\frac{n}{2}\right)}{\Gamma\left(\frac{n-1}{2}\right)} Sin^{n-2}(\theta) \end{aligned} \qquad (12)$$

This can be integrated with the limits from the observed principal angle to $\frac{\pi}{2}$ to derive the p-value.

*Evaluation of enrichment with the Hypergeometric test*

The lists of identified significant genes were examined for enrichment with various gene sets using the Hypergeometric test. The null hypothesis is that there is no enrichment and therefore the size of the overlap between the two gene lists should be distributed according to the Hypergeometric distribution:

$$p(k) = \frac{\binom{n_s}{k}\binom{n_s-n_t}{n-k}}{\binom{n_t}{n}} \qquad (13)$$

Where $p(k)$ is the probability of observing an overlap of size $k$; and $n$ is the length of the gene list; $n_s$ is the number of genes deemed significant according to the statistical test of differential expression, and $n$ is the total number of genes measured. The alternative hypothesis is that there is a significant enrichment with a p-value derived from the corresponding cumulative distribution. These p-values were converted into q-values with the Benjamini-Hochberg multiple hypothesis test and an FDR was set.

*Numerical simulations*

The approach we took in generating our synthetic data was to base it upon a highly simplified model which contains some of the broadest typical characteristics of real biological gene expression data. In a given biological state there is typically some random variability in the expression profile which can have technical as well as biological origin; we chose a probability distribution for our synthetic data which captures two typical basic properties of this random variation. Firstly, we require that a fraction of all the genes correlate with each other in their natural variation. Secondly, we require that the genome-wide variation has an intrinsic dimension which is quite low: 90% of the variation in genome-wide expression data can typically be captured in only a few principal components. We chose the simplest distribution which admits these two properties – the multivariate Gaussian distribution with mean vector $\mu$ and covariance matrix $\Sigma$. In order to minimize the number of free parameters, we set $\Sigma$ to be the same for both the control and the perturbed synthetic data and introduce differential expression via a difference in the mean vector for the control, $\mu_c$, and for the perturbation, $\mu_p$. We define a vector $S$ which is equal to the identity matrix with a number $d$ of diagonal entries replaced with a value $s$, where $d$ is the intrinsic dimensionality of the data. When $s \gg 1$ most of the variation in the data is contained in a $d$ dimensional subspace – this is how the low-dimensionality constraint is enforced. The matrix $S$ is then rotated randomly in an isotropic manner, into a subspace of dimension $c$ which is set according to the fraction of all genes which are correlating. This rotation is performed in terms of a random rotation matrix $R$ as follows:

$$\Sigma = RSR^T \qquad (14)$$

to derive the covariance matrix. The difference in the mean vectors defines the differential expression as $\Sigma$ is the same for both control and perturbed distributions. We simply generate an isotropic random vector in the correlating subspace described above. In order to control for the number of differentially expressed genes, we restrict the domain of this distribution to a subspace of dimension equal to the number of differentially expressed genes. All other components of $\mu_c - \mu_p$ are equal to zero. The random differential expression vector is scaled to set an overall magnitude of differential expression. Random sampling from the multivariate Gaussian distribution with the above described parameters is then used to generate synthetic data for control and perturbed states, with a known vector of differentially expressed genes.

# Results

We compare the Characteristic Direction approach with other popular methods for discovering differentially expressed genes as exemplified by the Welsh t test and other tests in a number of biological settings as well as in a numerical simulation.

*Validation of the methods on synthetic data through numerical simulations*

We first evaluate the Characteristic Direction method under a controlled situation where we generated synthetic expression data for numerical simulations. Real biological data has a structure which is far from being well defined but contain certain features that we can mimic; so we generated synthetic expression data as described in the methods. We then employed the t-test and our Characteristic Direction approach to test how these recover already known differentially expressed genes. We set the total dimensionality of expression space, i.e., the number of genes on the array, to 100; the intrinsic dimensionality to 2 $d = 2$; the variance parameter $s = 40$; the fraction of correlating and differentially expressed genes both to 0.1, and an overall magnitude of differential expression to 5.0. After 100 simulations the mean receiver operator characteristic (ROC) curve is computed (Fig. 2a). In order to further investigate the comparison between the t-test and the Characteristic Direction in recovering the known differentially expressed genes we examined the area under the ROC curve (AUC) transformed into the Gini Coefficient ($G = 2AUC - 1$), which is a measure of the efficiency of the approach in recovering the differentially expressed genes, as a function of the sample size (Fig. 2b).

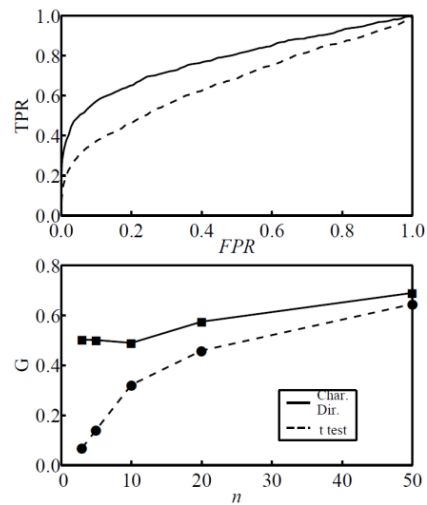

**Fig 2.** (a) The mean ROC curve for the recovery of differentially expressed genes over 100 numerical simulations with the following parameters: the total number of genes = 100, the intrinsic dimensionality of the data is 2, the fraction of differentially expressed genes and correlating genes is 0.1. (b) The efficiency of recovery of differentially expressed genes as measured by the Gini Coefficient, for the t-test and the Characteristic Direction approaches, averaged over 400 numerical simulations as a function of sample size. The parameters are the same as in (a) with the exception that the total number of genes is 50.

The results from these simulations suggest that the characteristic direction method is more sensitive that the t-test in recovering the differentially expressed genes and this is more pronounced when few samples are available for each condition, a typical situation in analyzing genome-wide gene expression data in biology.

*Validation of the methods with Estrogen receptor binding and differential expression*

Estrogens are natural hormone steroids that among other functions regulate cell growth and differentiation in the mammary gland and thus play a critical role in the development of breast cancer. Estrogens bind to nuclear receptor proteins that act as transcription factors in a ligand dependent manner. The nuclear receptors transcription factors estrogen receptor (ER) alpha and beta are two well-studied estrogen dependent factors that bind to DNA to transactivate the expression of estrogen-response target-genes. ER alpha and beta are structurally similar with similar binding sites. However, they serve different roles in the response to estrogens. These nuclear receptors can homo- or hetero-dimerize. It is known that ER beta has a lower transacting ability, and hetero-dimers also have a reduced efficiency in activating target genes than ER alpha homo-dimers. This difference could be explained by the recruitment of different co-factors, even though the DNA binding sites are similar. Malignant cells tend to have higher ER alpha and lower ER beta proportions than normal or benign tumor cells. To evaluate and validate our methods to extract the most important genes from expression data, we reanalyzed gene expression and ChIP-seq data collected from MCF7 breast cancer cells. The wild-type ER-beta+/ER-alpha+ MCF7 cells were compared to ER-beta-/ER-alpha+ engineered MCF7 cells after treatment with 17-beta-estradiol (E2) [17]. In their study, the authors first used ChIP-seq to identify the binding sites of ER-alpha and ER-beta in MCF7 cells. Sequence tags were aligned to the human genome and Model-Based Analysis of ChIP-Seq (MACS) was used to identify the peaks enriched after E2 treatment. The result was 9702 binding sites, or peaks, for ER alpha and 6024 for ER beta of which 4506 are shared. To identify the genes that are potentially regulated by these binding sites we computed the distance to the transcription start site (TSS) to the closest gene for each binding-site/peak. We then associate each binding site with a gene, and recorded the distance to the TSS of that gene. In addition to ChIP-seq, the authors of the study also measured gene expression after E2 stimulation of these cells. We reanalyzed the pre-processed Illumina expression array data to identify differentially expressed genes comparing the control to the two hour treatment sample profiles. Using the various methods described above, we next identified the Characteristic Directions for each of the methods as well as identified differentially expressed genes using the t-test. In the case of the Characteristic Direction approach, the most significant genes are defined by the magnitude of their component in the characteristic direction. This resulted in 640 (LR1), 640 (LR2), 640 (LDA), 657 (NP1), 633 (NP2) significant genes which collectively account for a total of 30% of the difference between the samples. The degree of significance for the t-test was chosen in order to ensure a sensible number of significant genes which was comparable to the numbers identified using the various Characteristic Direction approaches; this required an FDR threshold of 5%. 637 genes are differentially expressed at this level of significance with the t-test. The significant genes for LR1, LR2 and LDA are the same so further analysis used only LR1 which shall be referred to as the Characteristic Direction.

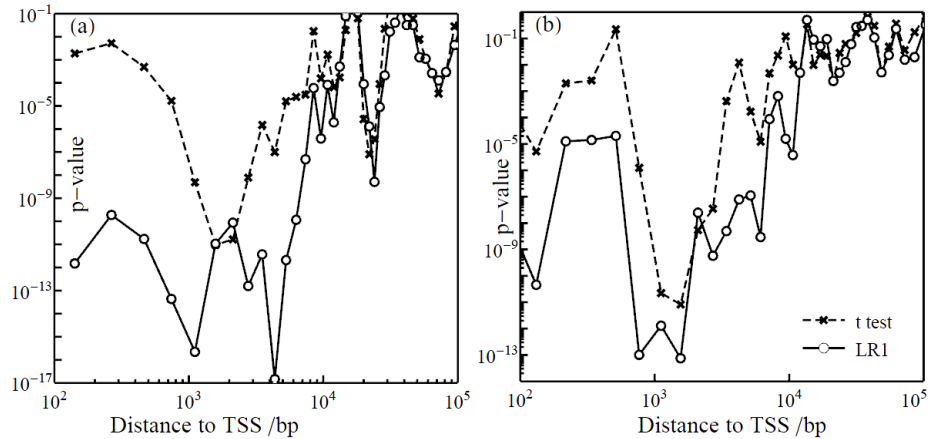

**Fig 3**. Enrichment of subsets of genes associated with ChIP-Seq binding sites in the gene expression data after two hour stimulation of MCF7 cells with a) ER alpha, and b) ER beta, plotted against their mean distance to the TSS of the nearest gene.

We next compared the ability of each method to identify putative ER direct target genes by comparing the significant genes from the various methods to the genes associated with binding sites identified from the Chip-seq data. We reason that the best performing method would identify a larger overlap between the ChIP-seq target associated genes and the significant genes identified in the differential expression. We take the list of ChIP-seq peaks ordered by the distance to the TSS of their nearest gene and remove duplicates by retaining only those genes associated with the maximally proximal binding site. Then using a sliding window of width 300 base pairs (bp) we plot the significance of enrichment against the mean distance to TSS of each subset of genes (Fig. 3). The results show that significant genes in the differential expression are significantly enriched for genes which are close to binding sites of ER. The degree of enrichment is particularly strong in the cases where the binding site is closest to the TSS. Superimposed on this overall trend are inflections in the curves that may point to enhancer and promoter regions. Most importantly, the Characteristic Direction method is always identifying more differentially expressed genes that are ER targets compared with the t-test. As ER alpha and beta are transcription factors which are known to transactivate primary response genes via DNA binding, this result suggests that the characteristic direction approach is more sensitive to identify differences in genome-wide expression profiles.

*Using a larger dataset of ChIP-seq and transcription factor perturbations from GEO, ENCODE and ChEA to validate the Characteristic Direction approach*

Next, we sought to expand the same type of validation process, which is matching binding sites for a transcription factor measured by ChIP-seq with differentially expressed genes as measured by microarrays after the transcription factor perturbation, on a larger dataset, comparing the Characteristic Direction method to the widely used t-test for these tow tests' ability of identifying the correct differentially expressed genes. We manually searched the Gene Expression Omnibus (GEO) [18] for studies that knocked-down or over-expressed transcription factors in mammalian cells. We also made sure that those perturbed factors also have DNA binding data in ENCODE [19] or ChEA [20]. We processed the ECONDE dataset into a gene-set library as previously described [21]; ChEA is already natively stored in a gene-set library. In total, we identified 16 experiments that constitute transcription factor perturbations that also have data in ENCODE, including 13 knockdowns and 3 over-expressions; and 47 experiments that also have entries in ChEA, including 37 knockdowns and 10 over-expressions.

We can now pick any one transcription factor perturbation experiment, for example, one of the experiments which knocked down HSF1, and examine the enrichment of the binding sites of HSF1 as determined by one of the experiments in ENCODE. Comparing the top n ranked genes as determined by the t-test or by the Characteristic Direction, we plot the number of overlapping genes (Fig. 4). We see that the top n genes from the Characteristic Direction approach always contain more overlapping HSF1 target genes compared with those called by the t-test.

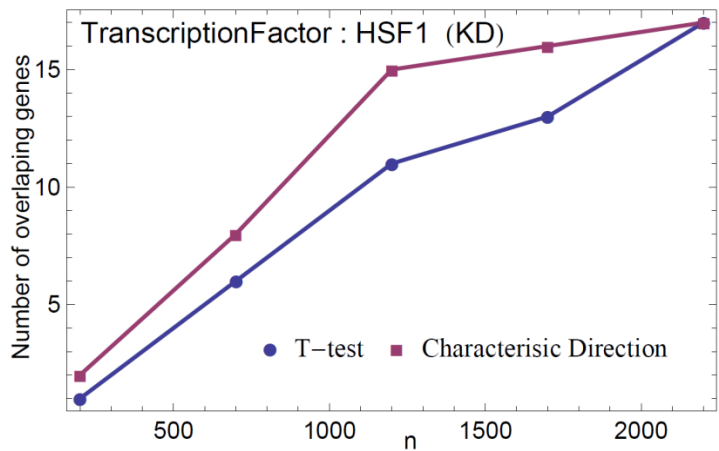

**Fig. 4.** The number of genes associated with the binding sites of HSF1 which overlap with the top n genes after HSF1 knockdown followed by gene expression testing the two ranked lists of differentially expressed genes as determined by the t-test or the Characteristic Direction.

And we can also examine the ratios between the numbers of overlapping genes as a function of n. While the above results for a single experiment HSF1 are suggestive of showing that the Characteristic Direction works better than the t-test, we could build a stronger case by examining all the experiments we found on GEO for which there are appropriate data. Below are plots of the mean ratio of the numbers of overlapping genes as a function of n using the knockdown experiments that match transcription factors in ENCODE (Fig. 5) and ChEA (Fig. 6), with a shaded area indicating one standard error. We observe that the values are always greater than one (~ 1.2), which indicates that there are about 20% more differentially expressed genes associated with the relevant transcription factor binding identified by the Characteristic Direction approach compared with the differentially expressed genes identified by the t-test. A similar result is also obtained for the over-expression experiments.

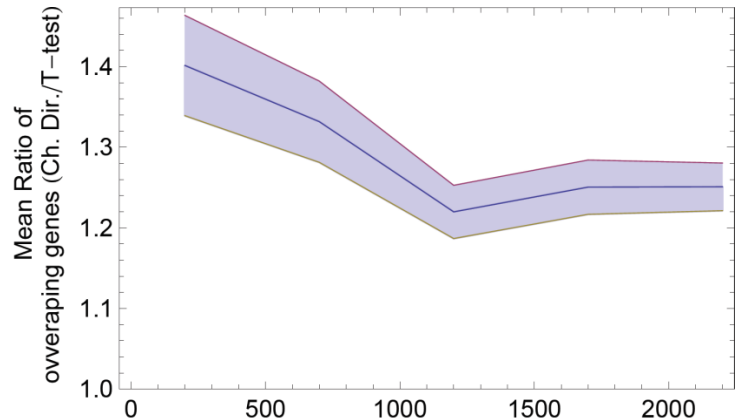

**Fig. 5**. Mean ratio of the number of overlapping (ENCODE) genes for the knock-down experiments. A value greater than unity indicates that the Characteristic Direction identifies more genes which are associated with binding sites of the knocked-down transcription factor compared with the differentially expressed genes identified by the t-test. The shaded area indicates one standard error.

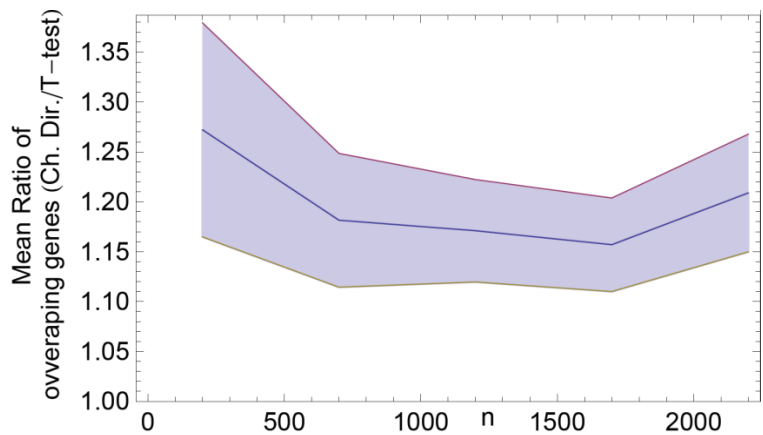

**Fig. 6**. Mean ratio of the number of overlapping (ChEA) genes for the knock-down experiments. A value greater than unity indicates that the Characteristic Direction identifies more genes which are associated with binding sites of the knocked-down transcription factor compared with the differentially expressed genes identified by the t-test. The shaded area indicates one standard error.

*Gene Ontology analysis of genome-wide breast cancer expression profiles*

The cancer stem cell model hypothesizes that tumors are organized in a cellular hierarchy in which a minority population of cells are responsible for tumor initiation, maintenance, recurrence and drug resistance. In breast cancer, a candidate for cancer stem cells has been identified as having the CD44+ CD24-/low surface markers. These cells, when injected into immune compromised mice, were shown to be highly tumorigenic, having the ability to invade and metastasize. In this case study we attempt to compare the various biological contexts that emerge when examining differentially expressed genes identified from mRNA profiling of the CD44+ CD24-/low breast cancer cells as compared with normal breast epithelium tissue. The data used in this case study for evaluation and validation comes primarily from a study that profiled and compared normal breast epithelium tissue obtained from reduction mammoplasties and highly tumorigenic breast cancer cells isolated from tumors (ESA+ CD44+ CD24-/low Lin-) [22]. The various approaches to identify differentially expressed genes produce different gene lists, or gene rankings, from this dataset; and this presumably provides different pictures of the biological mechanisms under investigation. When comparing CD44+ CD24-/low breast cancer stem cells with normal breast epithelium tissue we expect to detect biological processes such as cell motility, cell proliferation, wound healing [23], and extra cellular matrix (ECM) remodeling which are known to be up-regulated in cancer stem cells and are activated in aggressive tumors.

One commonly used approach to obtain a picture of the biology from the analysis of differential expression is the evaluation of the enrichment of gene sets. Gene Set Enrichment Analysis (GSEA) mentioned in the introduction, is one of the most popular approaches to accomplish this task. The more basic and widely used approach is to use a t-test to identify differentially expressed genes and then apply the Hypergeomtric test to examine enrichment of gene sets deriving from various gene-set libraries or the Gene Ontology [24]. We can use these methods, as well as the Characteristic Direction approach to evaluate and compare significant biologically meaningful gene sets. We first manually construct six subsets of GO biological processes corresponding to the six hallmark characteristic of cancer as defined by Hanahan and Weinberg: 1) regulation of cell proliferation; 2) evasion of growth suppression; 3) resisting cell death; 4) enabling replicative senescence; 5) induction of angiogenesis; and 6) enabling invasion and metastasis [25, 26]. In order to evaluate the significance of each of these gene-sets in the differential expression between the CD44+/CD24-/low samples and the normal breast epithelial samples [22] we first call the significant genes with the t-test and the two methods of calculating the Characteristic Direction LR1 and NP1 before evaluating enrichment.

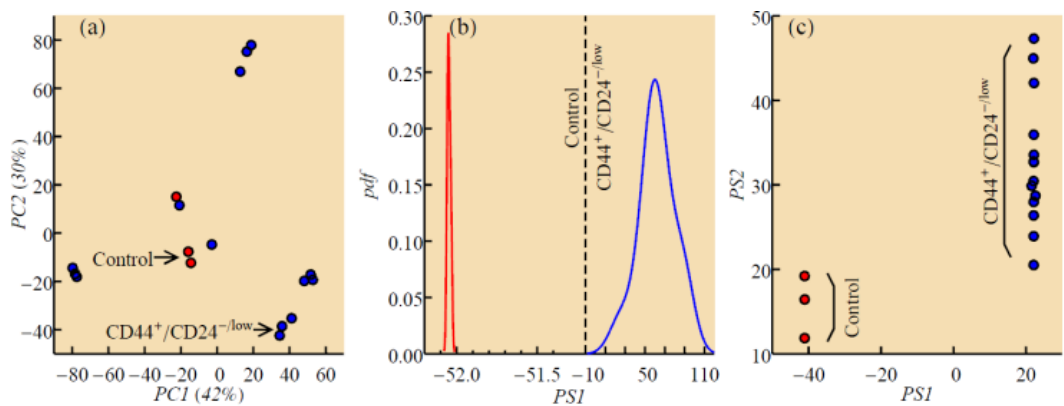

**Fig 7.** Three different visualizations of the CD44+/CD24-/low verses normal breast epithelium data reveal that the Characteristic Direction approach efficiently separates the samples and captures their differential expression. (a) A PCA plot of the expression data shows no clear separation of the normal breast samples (red) and the tumorigenic samples (blue) as the directions identified in this approach (principal components) are dominated by the high variance in the tumorigenic samples. (b) The same data is projected onto the characteristic direction. The red and blue curves are Gaussian kernel density estimates of the distributions of the normal and tumorigenic distributions respectively. The dashed line indicates a break in the axis separating the two distributions for the purpose of clarity. (c) An extension of the plot shown in (b) into two characteristic dimensions (CD1 and CD2) as described in the methods.

The projection of the breast cancer data onto the direction that best characterizes the differences between the normal breast epithemium and the CD44+/CD24-/low cancer stem cell samples can be visualized as described in the Methods (Fig. 7). This visualization shows that the PCA plot shows little or no difference between the tumorigenic and normal samples. However, in contrast, when the data is projected onto the characteristic direction chosen by the approach described above (Fig. 7b), the result is a clear separation of the samples, which indicates that the direction derived, effectively characterizes the difference between these two biological conditions. This visualization approach is taken a step further in Fig. 7c where it is extended to two dimensions. Next, we compared the six different approaches for evaluating the significance of the above GO categories in the differential expression between the tumorigenic CD44+/CD24-/low samples and the normal breast epithelium samples. The first three of these are similar in that they first identify a list of genes which are significant in the differential expression before using the Hypergeometric test to evaluate the enrichment of the GO categories and correcting for multiple hypotheses testing with the Benjamini-Hochberg test. In the case of the t-test, a threshold FDR of 1% results in 657 significantly differentially expressed genes. In the case of the Characteristic Direction with linear regression, the subset of the most significant genes, which collectively account for 50% of the total magnitude of differential expression, constitutes 655 genes. In the case of the non-parametric method the number is 797. The threshold levels were chosen to be of reasonable significance while also resulting in similar size gene lists for appropriate comparison. The differentially expressed genes determined by the t-test could be divided into up-regulated and down-regulated genes, and the significant genes by the Characteristic Direction could be divided into positively and negatively discriminant. We observe no overlap between the up-regulated genes and the negatively discriminant genes, and between the down-regulated genes and the positively discriminant genes, as expected. The overlap between the up-regulated genes and the positively discriminant genes in the case of the linear regression approach is 63, and for the down-regulated genes and the negatively discriminant genes is 85 which is indicative of a limited degree of similarity between the two approaches. Two further approaches to the evaluation of the enrichment of the GO category gene sets were taken; one in which the principal angle (which is related to the canonical correlations) between the two characteristic directions LR1 and NP1 and the subspace of expression space induced by the gene set is used (see Methods). These two tests are referred to as PA1 and PA2 respectively. The sixth and final approach to the enrichment via the GSEA online tool [27]. This referred to simply as GSEA.

Having evaluated the significance of each of the manually compiled GO categories, we compared the results from each of the above six approaches (Fig. 8). Colored squares indicate instances in which the corresponding method has found the corresponding GO category significant at an FDR of 10%. Squares are colored according to the mean rank of the overlaps with deep red indicating significant up-regulation and deep blue indicating significant down regulation, while paler colors indicating a mixture.

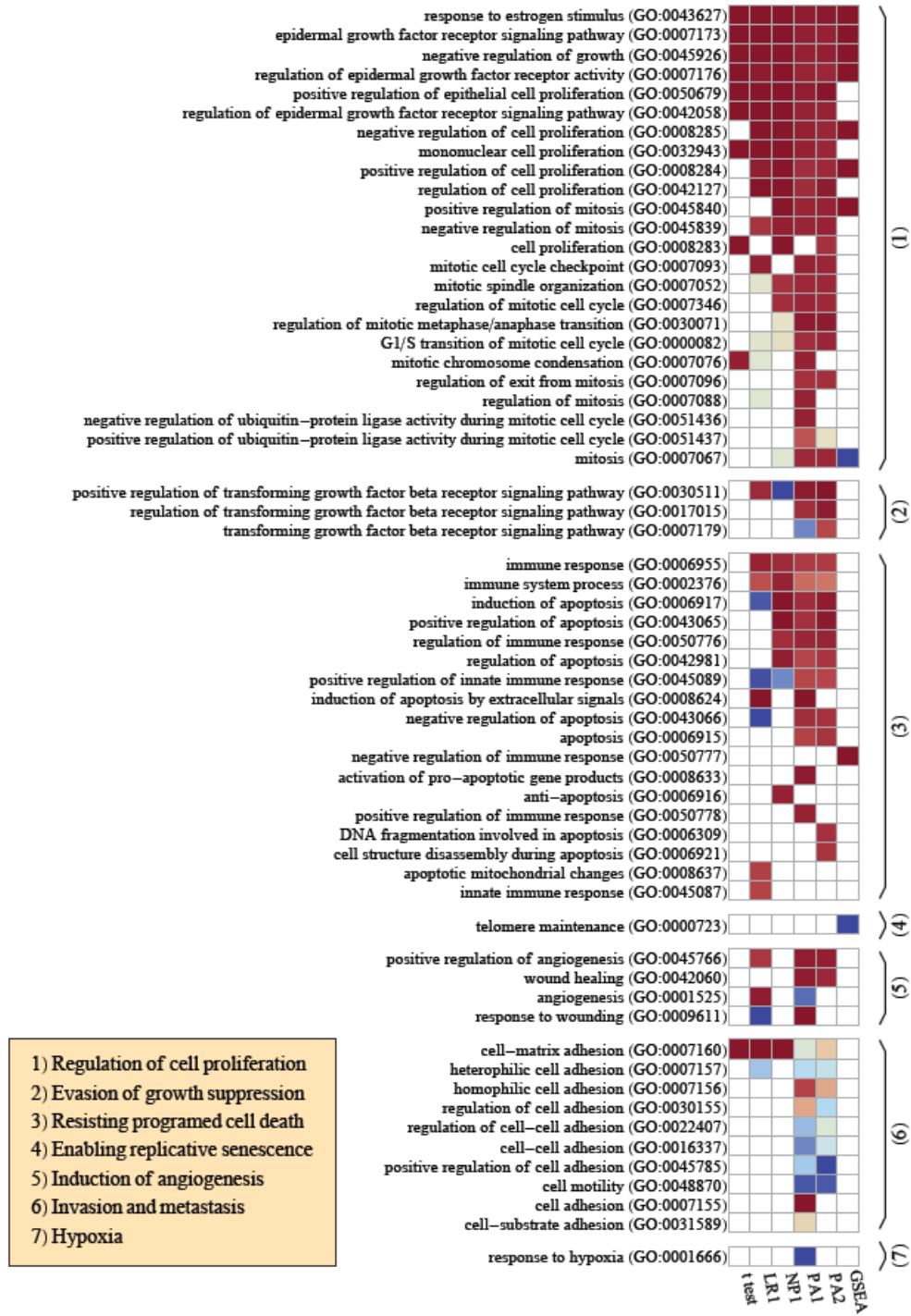

1) Regulation of cell proliferation
2) Evasion of growth suppression
3) Resisting programed cell death
4) Enabling replicative senescence
5) Induction of angiogenesis
6) Invasion and metastasis
7) Hypoxia

**Fig 8.** A comparison of GO categories identified as significant in the differential expression of tumorigenic verses normal samples by enrichment of the significant genes called by the t-test, and the Characteristic Direction methods LR1 and NP1, and also using the principal angle approach in conjunction with these directions, PA1 and PA2 respectively. We also include the results of GSEA [9] for comparison. Colored boxes indicate that the GO category is identified as significant with an FDR of 10%, and deeper red colors have a smaller mean rank of the gene set, corresponding to more up-regulation of the set, while deeper blue colors have a larger mean rank, corresponding to more down-regulation of the set. The GO categories are sub-categorized corresponding to six hallmark characteristics of cancer as indicated in the inset box. The seventh category is included to evaluate the significance of the hypoxia GO category.

We observe complete agreement between all the methods in GO categories concerned with mitosis, a central process in cell proliferation which is widely regarded as a fundamental biological process in cancer. The GO category of cell proliferation (GO0008283) is only found to be significant in the Characteristic Direction approaches and GSEA. We also note that all the processes identified as significant by GSEA (ten in total), with the exception of two, are also identified by the Characteristic Direction approaches. Overall it appears that the Characteristic Direction approaches find more of the cancer hallmark associated processes to be significant in the differential expression between the highly tumorigenic cells and the normal cells, which could possibly indicate that the Characteristic Direction method is more sensitive in identifying more relevant biological terms.

**Conclusions**

Global genome-wide differential expression of mRNAs, proteins and other types of biomolecules are continually accumulating at an accelerating pace. Although the cost of such experiments continually drops, such experiments are still expensive. Hence, extracting maximal knowledge from of such data is critical. Here we demonstrate that the commonly used t-test approach to extract differentially expressed genes could be improved by an alternative geometrical approach which results in a cleaner signal. To demonstrate the advantage of this Characteristic Direction approach we developed numerical simulations and several case studies applied to real data. In the first case study we used the binding sites of ER alpha and beta, as measured by ChIP-seq experiments, to identify genes whose expression is likely to be affected by the treatment of modified MCF7 cells with E2, an ER stimulating ligand. The identified genes were then used to validate our approach by demonstrating that they are significantly more enriched, and thereby identified as significantly affected by treatment with E2, than observed with differentially expressed genes as suggested by the t-test. We also observed some interesting features in the dependence of the enrichment upon distance from the TSS of the nearest gene. In a second case study, we expanded this analysis to many other similar studies that either knocked-down or over-expressed transcription factors that also have published data from ChIP-seq profiling of the factors. Our results show that the Characteristic Direction approach consistently recovers over 20% more of the relevant target genes of the perturbed factor compared with genes recovered by the t-test. To further demonstrate that the Characteristic Direction approach outperforms other tests, in the final case study we examined enriched GO terms comparing breast cancer stem cells to normal breast epithelium. The Characteristic Direction approach is able to identify more relevant genes that uncover more relevant biological categories in a more robust way. These few case studies show that the Characteristic Direction method is superior to other popular methods of analysis of differential expression data and suggest that the Characteristic Direction method should be applied in the future in studies that perform genome-wide expression profiling. To make the approach accessible, we implemented the method in Python, MATLAB and Mathematica. Readers that are interested in applying the method to their own data should refer to the open source scripts and examples available at: http://www.maayanlab.net/CD.

## Acknowledgements

*Funding*: This work was supported in part by grants from the NIH: R01GM098316-01A1, P50GM071558, R01DK088541-01A1, U54HG006097-02S1 and RC4DK090860-01.